\documentclass[12pt,a4paper,DIV12]{scrartcl}
\usepackage[utf8]{inputenc}     
\usepackage[T1]{fontenc}        
\usepackage{lmodern}            
\usepackage[british]{babel}
\usepackage{amsmath}
\usepackage{amssymb}
\usepackage{amsfonts}
\usepackage{setspace}           
\usepackage{graphicx}
\usepackage{hyperref}
\usepackage{color}
\title{%
  {\vspace{-20mm} \normalsize
   \hfill\parbox[b][30mm][t]{35mm}{\textmd{MS-TP-20-20}}}\\[-18mm]
Interface roughening in two dimensions}
\author{Gernot M\"unster%
\thanks{munsteg@uni-muenster.de}
\ and
Manuel Ca\~{n}izares Guerrero%
\thanks{manolok96@gmail.com, present address:
Faculty of Sciences, Universidad de Granada,
Fuente Nueva s/n, 18071 Granada, Spain}\\[2ex]
University of M\"unster, Institute for Theoretical Physics,\\
Wilhelm-Klemm-Str.~9, D-48149 M\"unster, Germany}
\date{April 28, 2020}

\begin{document}
\maketitle

\begin{abstract}
Roughening of interfaces implies the divergence of
the interface width $w$ with the system size $L$. For two-dimensional
systems the divergence of $w^2$ is linear in $L$. In the framework of
a detailed capillary wave approximation and of statistical field theory
we derive an expression for the asymptotic behaviour of $w^2$, which differs
from results in the literature. It is confirmed by Monte Carlo simulations.
\end{abstract}
\vspace{5mm}
%
\section{Introduction}

Roughening of interfaces in three-dimensional thermodynamical systems
is a well-studied phenomenon 
\cite{Mandelstam:1913,Burton:1949,BLS:1965,Weeks:1973,Weeks:1977,Rowlinson:1982,Jasnow:1984}. 
A system may under suitable conditions
show coexistence of two phases that are spatially separated by an
interface. In a range of temperatures $T_R < T < T_c$ below the critical
point, the corresponding breaking 
of translation invariance is associated with Goldstone modes, which 
manifest themselves as long-wavelength fluctuations of the interface 
and lead to its roughening.
The most characteristic feature of interface roughening
is the fact that the width of the interface increases with the size 
of the system.
In the so-called capillary wave approximation \cite{Mandelstam:1913,BLS:1965}
the width $w$ of an
interface with quadratic shape $L \times L$ is given by an integral 
over wave numbers $q$
\begin{equation}
\label{eq:cwd3}
w^2 = \frac{1}{\sigma} \int\! \frac{d^2 q}{(2\pi)^2}\, \frac{1}{q^2}
= \frac{1}{2 \pi \sigma} \int_{q_{\textrm{min}}}^{q_{\textrm{max}}} 
\frac{dq}{q}
\qquad (d=3),
\end{equation}
where $\sigma = \tau /kT$ is the reduced interface tension.
The lower cutoff on wave numbers is given by $q_{\textrm{min}} = 2 \pi/L$.
The upper cutoff represents the scale beyond which the capillary wave
approximation ceases to be valid. Assuming that the lower limit on the
wavelengths of capillary waves is of the order of the correlation length $\xi$,
we write $q_{\textrm{max}} = 1/c\,\xi$ with an adjustable numerical factor $c$.
This leads to
\begin{equation}
\label{eq:w2d3}
w^2 = \frac{1}{2 \pi \sigma} \ln \frac{L}{2 \pi c\, \xi}
\qquad (d=3).
\end{equation}
The special feature of this expression is the fact that the logarithmic
divergence and the prefactor
$1/2 \pi \sigma$ are universal and do not depend on the details of the
cutoffs. In particular, the expression is not affected by the fact that the proper
discrete sum over wave numbers is approximated by an integral.
It also does not depend on the microscopic structure of the model.
In fact, a complete exact calculation in renormalised three-dimensional field theory
to one-loop order without any ad hoc cutoffs \cite{Koepf:2008} confirms that the
large-$L$ behaviour of the interface width is given by Eq.~(\ref{eq:w2d3}).
It has also been observed in a number of Monte Carlo investigations
of interfaces in the three-dimensional Ising model, 
see \cite{Werner:1999,Muller:2004vv} and references therein.

In two dimensions the situation is completely different. A direct transcription
of the capillary wave formula (\ref{eq:cwd3}) to two dimensions gives
\begin{equation}
\label{eq:cwd2}
w^2 = \frac{1}{\sigma} \int\! \frac{dq}{2\pi}\, \frac{1}{q^2}
= \frac{1}{\pi \sigma} \int_{q_{\textrm{min}}}^{q_{\textrm{max}}} 
\frac{dq}{q^2}
= \frac{L}{2 \pi^2 \sigma} - \frac{c \,\xi}{\pi \sigma}
\qquad (d=2) .
\end{equation}
Here the expression for the leading linear divergence with $L$ is not
universal, depending on the summation of the wave numbers as an integral 
and on the particular choice of the lower cutoff. 

Before going into more details let us make some general remarks about
interface roughening in two dimensions. In this case the ``interface'',
separating coexisting phases, is a one-dimensional line.
The most prominent two-dimensional model, in which interfaces have been studied,
is the Ising model, but other models have been studied as well with comparable
results. To be definite, we shall consider the Ising model 
on an $L \times M$ square lattice with unit lattice spacing in the following.
In the low temperature phase, an interface along the $L$-direction
can be forced to be present by
suitable boundary conditions. One possibility is to choose fixed positive spins
on half of the boundary and negative spins on the other half. Another choice
are periodic boundary conditions in the $L$-direction, and antiperiodic ones in the
$M$-direction. These boundary conditions have the advantage to respect
translational invariance of the system.
There exist different definitions for the interface tension \cite{Abraham:1986}, 
which, however, lead to the same value \cite{Onsager:1944}
\begin{equation}
\sigma = 2 \beta + \ln \tanh \beta \qquad (\beta > \beta_c),
\end{equation}
where $\beta = 1/kT$ is the inverse temperature. In addition, the interface is
related to the correlation length $\xi$ by the exact identity
\begin{equation}
2 \sigma \xi = 1\,. 
\end{equation}

It has been shown rigorously that roughening is present
for all temperatures below the critical temperature
and that the leading divergence of the squared interface width $w^2$ 
with the system extent $L$ is 
linear \cite{Gallavotti:1972,Abraham:1974,Abraham:1986}. 
For the form of this linear term one finds, however, different
results in the literature. Gelfand and Fisher \cite{Gelfand:1990} discuss 
the capillary wave approximation with
\begin{equation}
\label{eq:GF}
w^2 \sim \frac{L}{2 \pi^2 \sigma}
\end{equation}
as in Eq.~(\ref{eq:cwd2}). Based on exact results for the Ising model,
Abraham \cite{Abraham:1981} and Fisher et al.~\cite{Fisher:1982} give
\begin{equation}
\label{eq:AF}
w^2 \sim \frac{L}{\Gamma},
\end{equation}
where
\begin{equation}
\Gamma = \sinh \sigma
\end{equation}
can be interpreted as an effective interface tension in a generalised
capillary wave approximation.

%
\section{Theoretical Results for the Interface Width}

In view of these conflicting results we shall investigate the divergence
of the interface width with the system size $L$ more thoroughly.
In the capillary wave approximation the interface is in the scaling region
near the critical point represented by a continuous line, whose transverse
elongation $h(x)$ (``height'') is Gaussian distributed with a probability
distribution
\begin{equation}
p_{\text{CWA}} \propto 
\exp \left\{ - \frac{\sigma}{2} \int\!dx\, 
\left( \frac{dh}{dx} \right)^2 \right\}.
\end{equation}
In a system of finite extent 
$L$ the allowed wave numbers for the height function are
\begin{equation}
q_n = \frac{2 \pi}{L} n, \quad \text{with} \quad n \in \mathbf{Z},
\end{equation}
and their increment is $\Delta q = \frac{2 \pi}{L}$.
The interface width $w$ is given by
\begin{equation}
\label{eq:cwa}
w^2 = \langle h^2 \rangle 
= \frac{1}{\sigma} \sum_{q_n \neq 0} \frac{\Delta q}{2 \pi} \frac{1}{q^2}.
\end{equation}
The zero mode $q_0 = 0$ describes a rigid translation of the interface and has to
be omitted in the sum. Furthermore we may introduce an upper cutoff of the order
of the inverse correlation length
\begin{equation}
q_{\text{max}} = \frac{2 \pi}{L} N
\doteq 1 / c \,\xi ,
\end{equation}
leading to
\begin{equation}
w^2 = \frac{2}{\sigma L} \left( \frac{L}{2 \pi} \right)^2 \,
\sum_{n=1}^{N} \frac{1}{n^2} .
\end{equation}
We write the sum as
\begin{equation}
\sum_{n=1}^{N} \frac{1}{n^2} =
\sum_{n=1}^{\infty} \frac{1}{n^2} -
\sum_{n = N + 1}^{\infty} \frac{1}{n^2} .
\end{equation}
The first term is given by Euler's solution of the Basel 
problem \cite{Euler:1740}:
\begin{equation}
\sum_{n=1}^{\infty} \frac{1}{n^2} = \frac{\pi^2}{6} .
\end{equation}
The second term can be estimated by Euler's summation formula \cite{Euler:1738}:
\begin{align}
\sum_{n = N + 1}^{\infty} \frac{1}{n^2}
&= \sum_{n = N}^{\infty} \frac{1}{n^2} - \frac{1}{N^2}
= \int_{N}^{\infty} dx\, \frac{1}{x^2}
+ \frac{1}{2} \frac{1}{N^2}
- \frac{1}{N^2}
+ O(N^{-3}) \\
&= \frac{1}{N} - \frac{1}{2} \frac{1}{N^2} + O(N^{-3}) .
\end{align}
Putting things together we arrive at
\begin{equation}
\label{eq:w2}
w^2 = \frac{1}{12 \sigma} L
- \frac{c \,\xi}{\pi \sigma}
+ O(1/L) .
\end{equation}
The leading large-$L$ term is clearly different from the cited predictions in
Eqs.~(\ref{eq:GF}),(\ref{eq:AF}). Its form depends on the specific infrared 
cutoff by taking the discreteness of the wave numbers into account.

This result can be confirmed by statistical field theory. In the scaling region
the microscopic degrees of freedom are described by a real valued order parameter
field $\phi(x)$, governed by a Landau-Ginzburg Hamiltonian \cite{Bellac:1991}. 
In the two-phase region the interface is represented by an interface profile 
function $\phi_R(x)$, given by a classical solution \cite{vdW:1893,Cahn:1958}
plus fluctuation contributions \cite{Jasnow:1978,Muenster:1990}.
The fluctuation contributions in three dimensions have been calculated analytically 
in one-loop order by means of the inverse fluctuation operator in the interface 
background field \cite{Koepf:2008}.
The resulting profile function deviates from the 
Fisk-Widom scaling form \cite{Fisk:1960} by depending logarithmically
on the system size $L$. The corresponding calculation can be performed for the
two-dimensional case. It is not appropriate to repeat the details of the method here.
We restrict ourselves to point out that the
leading large-$L$ term of the profile function, given in Eqs.~(50),(52) of
Ref.~\cite{Koepf:2008} for the three-dimensional case, contains a contribution with a 
factor $C_1$, 
which in two dimensions reads
\begin{equation}
C_1 = \frac{3 L}{32 \pi^2 \xi} \sum_{n \neq 0} \frac{1}{n^2}
= \frac{3 L}{16 \pi^2 \xi} \sum_{n=1}^{\infty} \frac{1}{n^2}
= \frac{L}{32\, \xi} .
\end{equation}
The interface width is defined through the second moment of a distribution $p(x)$,
\begin{equation}
w^2 = \int\!dx\, x^2 p(x),
\end{equation}
which is defined through the derivative of the profile as in Ref.~\cite{BLS:1965},
\begin{equation}
p(x) \propto \partial_x \phi_R(x), \qquad \int\!dx\, p(x) = 1.
\end{equation}
The squared width $w^2$ has an $L$-dependent contribution
\begin{equation}
\frac{8}{3 \sigma} \xi C_1
= \frac{1}{12 \sigma} L
\end{equation}
in exact agreement with the result (\ref{eq:w2}).

The capillary wave approximation (\ref{eq:cwa}) can even be extended to
take the discreteness of the lattice into account.
If there are $L$ lattice points along the interface, the allowed
wave numbers for periodic boundary conditions are given by
\begin{equation}
q_n = \frac{2 \pi}{L} n, \quad \text{with} \quad n = 0, 1, 2, \ldots, L-1.
\end{equation}
The inverse Laplacean in wave number space is given by \cite{MM}
\begin{equation}
\frac{1}{\hat{q}_n^2} \quad \text{with} \quad
\hat{q}_n = 2 \sin \frac{q_n}{2} .
\end{equation}
The squared interface width is then given by
\begin{equation}
w^2 = \frac{1}{\sigma L} \sum_{n=1}^{L-1} \frac{1}{\hat{q}_n^2} 
= \frac{1}{4 \sigma L} \sum_{n=1}^{L-1} \frac{1}{\sin^2 \frac{\pi n}{L}} .
\end{equation}
For large $L$ the sum can be estimated with the help of Euler's summation formula.
Leaving out the details, the result is
\begin{equation}
\sum_{n=1}^{L-1} \frac{1}{\sin^2 \frac{\pi n}{L}}
= \frac{L^2}{3} + \text{const.} + O(1/L),
\end{equation}
and consequently
\begin{equation}
w^2 = \frac{1}{12 \sigma} L + O(1/L) .
\end{equation}

%
\section{Monte Carlo Simulation}

\begin{figure}[hb!]
\centering
\includegraphics[width=0.8\textwidth]{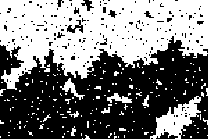}
\caption{A configuration of Ising spins on a $150 \times 100$ lattice
showing an interface.}
\label{fig:conf}
\end{figure}

In order to check the theoretical result we have performed a Monte Carlo 
simulation of the two-dimensional Ising model on a $L \times M$ lattice with
periodic/antiperiodic boundary conditions. The lattice size has been chosen
to be $M = 100$ and $40 \leq L \leq 200$ in steps of 10. Configurations have
been generated with the Metropolis method at an inverse temperature
$\beta = 0.46905$ slightly above
$\beta_c = \frac{1}{2} \ln (1 + \sqrt{2}) = 0.44069$.
The corresponding correlation length and interface tension are
$\xi = 4.49$ and $\sigma = 0.111$.
Due to the antiperiodic boundary conditions
in one direction, an interface forms in the system. An example configuration
is shown in Fig.~\ref{fig:conf}.

At each value of $L$ a number of thermalisation sweeps
has been performed, from 5.000 at $L=40$ to 20.000 at $L=200$,
followed by sweeps for the measurements. The total number of sweeps has been
700.000 for each $L$.
The measured integrated autocorrelation time varies between 20 to 30 sweeps.
Statistical errors have been estimated by means of the Jackknife method.

\begin{figure}[hb!]
\centering
\includegraphics[width=\textwidth]{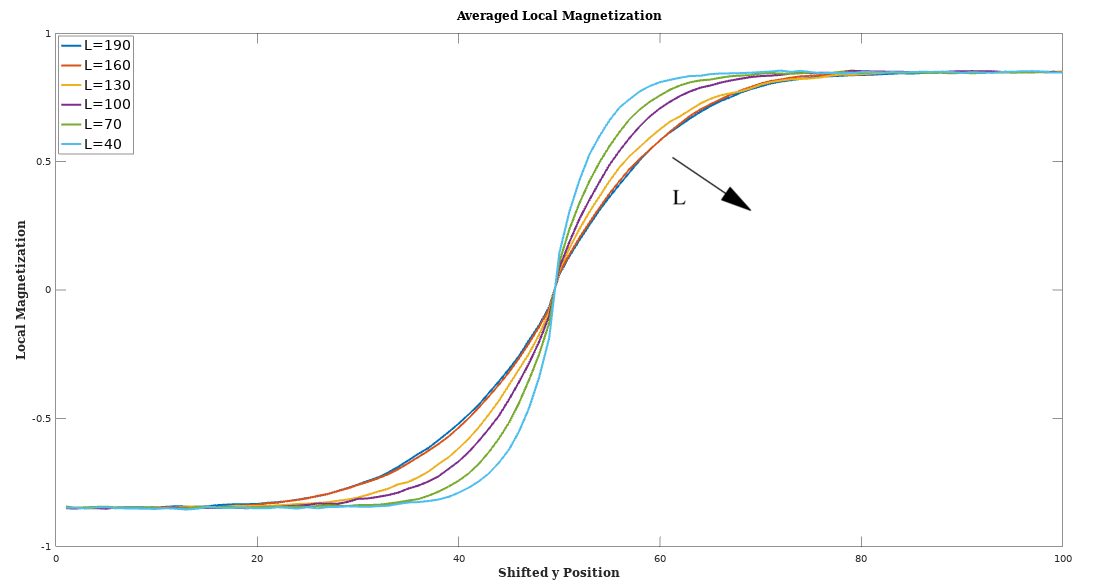}
\caption{Numerical results for the interface profiles for some values of $L$.
The broadening of the interface with increasing $L$ is clearly visible.
For better visibility the numerical errors are not displayed.
The errors of the widths are shown in Fig.~\ref{fig:widths}.}
\label{fig:profiles}
\end{figure}

For each configuration the interface is being centered, respecting translational
invariance, by means of the 
boundary shift method, which is explained in detail in Ref.~\cite{Muller:2004vv}.
The interface profiles are then calculated as the statistical averages of the
mean magnetisation in columns of length $L$. For some values of $L$ the resulting
profiles are displayed in Fig.~\ref{fig:profiles}.

\begin{figure}[htb]
\centering
\includegraphics[width=\textwidth]{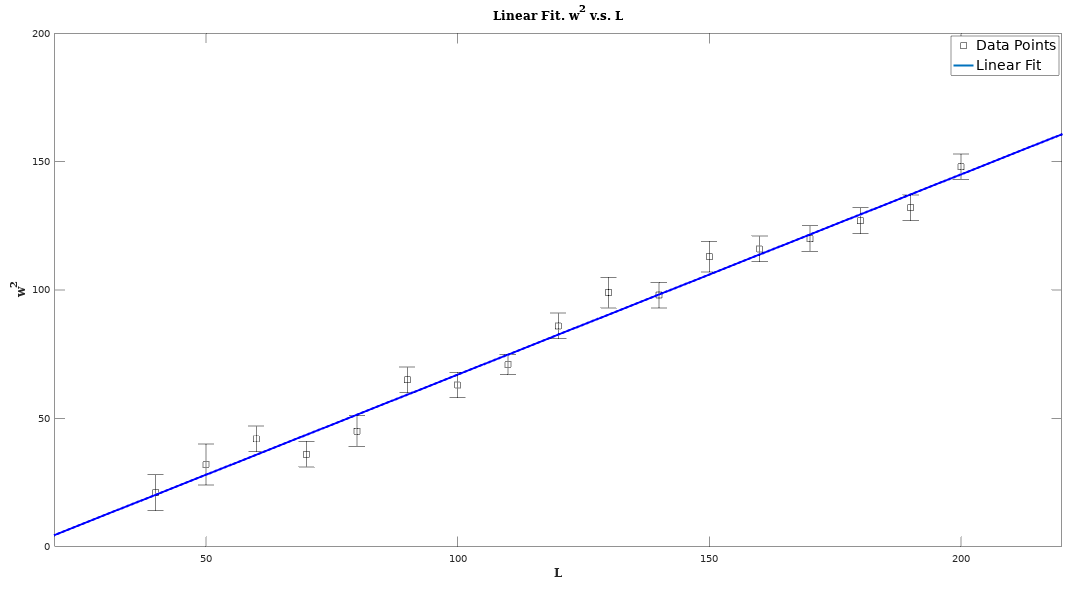}
\caption{Squared interface widths plotted versus system size $L$,
and a linear fit to the data.}
\label{fig:widths}
\end{figure}

For each obtained interface profile its squared width $w^2$ is calculated as 
the second moment of the distribution defined through the normalised derivative 
of the profile. Fig.~\ref{fig:widths} shows the results as a function of $L$
together with a linear fit of the form $w^2 = a L + b$. For the coefficients we find
$a = 0.78 \pm 0.03$ and $b = - 11 \pm 4$.
With the present values for $\sigma$ and $\xi$, the theoretical result (\ref{eq:w2}) 
amounts to
\begin{equation}
w^2 = 0.749\, L - 12.9\, c .
\end{equation}
The slope $a = 0.78 \pm 0.03$ is well consistent with the predicted 
value, confirming the theoretical result.
The coefficient $c$ in the intercept is consistent with a value near 1.

%
%
\section{Conclusion}

The broadening of interfaces is characteristic for the roughening
phenomenon. For two-dimensional systems the squared interface width
diverges linearly with the system size. In contrast to the three-dimensional
case, the precise form of this behaviour is not universal and does depend on
the details of the model. In the framework of a detailed capillary wave
approximation and of statistical field theory we have obtained a theoretical
prediction for the divergence of the interface width. The result is confirmed
by means of numerical simulations of the two-dimensional Ising model.

%
\end{document}